\documentclass[10pt]{article}
\usepackage[utf8]{inputenc}

\usepackage[letterpaper]{geometry}
\usepackage{hicss}
\usepackage{times}
\usepackage[none]{hyphenat}
\usepackage{url}
\usepackage{latexsym}
\usepackage{minted}
\usepackage{indentfirst}
\usepackage{graphicx}
\usepackage{tabularx}
\graphicspath{{images/}}
\usepackage[
    style=apa,
  ]{biblatex}
\title{Towards Effective E-Participation of Citizens in the European Union: \linebreak The Development of AskThePublic}
\author{}
\author{Nils Messerschmidt \\
SnT, University of Luxembourg \\
{\underline{nils.messerschmidt@uni.lu}} \\ \\
Xiaohui Wu \\
University of Zurich \\
{\underline{xiaohui.wu@uzh.ch} } \\ \And
Kilian Sprenkamp \\
University of Zurich \\
{\underline{kilian.sprenkamp@uzh.ch} } \\ \\
Igor Tchappi\\
SnT, University of Luxembourg  \\
{\underline{igor.tchappi@uni.lu} } \\ \\
Gilbert Fridgen\\
SnT, University of Luxembourg \\
{\underline{gilbert.fridgen@uni.lu} } \\ \And
Amir Sartipi \\
SnT, University of Luxembourg \\
{\underline{amir.sartipi@uni.lu} } \\ \\
Liudmila Zavolokina\\
University of Lausanne \\
{\underline{liudmila.zavolokina@unil.ch} } \\ }

\date{}

\begin{document}
\maketitle
\begin{abstract}
E-participation platforms are an important asset for governments in increasing trust and fostering democratic societies. By engaging public and private institutions and individuals, policymakers can make informed and inclusive decisions. However, current approaches of primarily static nature struggle to integrate citizen feedback effectively. Drawing on the Media Richness Theory and applying the Design Science Research method, we explore how a chatbot can address these shortcomings to improve the decision-making abilities for primary stakeholders of e-participation platforms. Leveraging the \textit{"Have Your Say"} platform, which solicits feedback on initiatives and regulations by the European Commission, a Large Language Model-based chatbot, called \textit{AskThePublic} is created, providing policymakers, journalists, researchers, and interested citizens with a convenient channel to explore and engage with citizen input. Evaluating \textit{AskThePublic} in 11 semi-structured interviews with public sector-affiliated experts, we find that the interviewees value the interactive and structured responses as well as enhanced language capabilities.
\end{abstract}

\subsubsection*{Keywords:}

E-Participation, Citizen Involvement, Large Language Models, Retrieval-Augmented Generation, Media-Richness-Theory.

\section{Introduction}

Fueled by a decade of crises since the global financial crisis of 2007-2008, including the Euro crisis, the refugee crisis, and the COVID-19 pandemic, polarization is on the rise across the European continent \parencite{matthijs2020lessons}. The increasing support for Euroscepticism and anti-political establishment parties is a particular concern for the bloc in major European countries such as France, Germany, and Italy, who are the largest net contributors to the European Union (EU) budget \parencite{begg2017eu} and whose citizens have mostly enjoyed the benefits of European cooperation since the 1950s. Meanwhile, the Brexit movement has led the UK to withdraw from the EU entirely. A common theme for anti-establishment parties is populist, anti-system or protest tendencies, and, in fewer cases, even radical or extremist ones \parencite{casal2021polarization}. 

One proposed way to increase trust in governments and thus decrease anti-systemic tendencies is citizen participation \parencite{kim2012participation}. More recently, governmental agencies have been promoting e-participation tools, such as the \textit{Have Your Say}\footnote{\url{https://ec.europa.eu/info/law/better-regulation/have-your-say}} platform, which invites its users to provide feedback to ongoing initiatives led by the European Commission. Despite successfully receiving citizen feedback, the platform lacks effective ways to structure, analyze and make meaningful conclusions out of this feedback, thus decreasing the ability to leverage the platform for primary stakeholders. These include EU policymakers, who aim to incorporate various perspectives for inclusive decision-making; researchers and journalists, who can leverage the provided feedback for scientific or popular research, and lastly, interested EU citizens, who can use the platform for information purposes.

Addressing the platform's shortcomings, the Media Richness Theory (MRT) offers a theoretical explanation for the various degrees of effective communication depending on the chosen medium. In accordance with the theory, a medium is more effective, the more interactive it is (\citeauthor*{daft1984information}, 1984; 1986). Assuming this theoretical lens, this would call for more interactivity on the \textit{Have Your Say} platform to be of more effective use to its stakeholders, in line with the concept of "conversational governments" \parencite{baldauf2020towards}. One method to increase interactivity while simultaneously avoiding the common issue of higher cost associated with higher media richness \parencite{androutsopoulou2019transforming}, e.g., face-to-face interactions, is the use of Large Language Models (LLMs), which have gained increasing public interest since the release of ChatGPT in 2022. This poses the following research question (RQ): 

\textit{How can an LLM-based chatbot leveraging e-participation data be designed to increase the informedness of primary stakeholders involved in EU policy-making processes?}

Our project resulted in the development of a proof-of-concept prototype based on the Design Science Research (DSR) methodology \parencite{hevner2004design, peffers2007design}, i.e., an LLM-enabled chatbot named \textit{AskThePublic}, designed to enhance e-participation mechanisms by facilitating the effective feedback analysis of its primary stakeholders.

\section{Related Work}
\subsection{Citizen Involvement in the EU}
Since the 1960s, governments have increasingly recognized public participation as a cornerstone in improving the legitimacy, accountability, and transparency of decision-making \parencite{lourencco2007incorporating}. Since then, an increasing number of legislative systems have solidified the right to public participation in law, recognizing it as a fundamental right to contribute meaningfully to decision-making processes \parencite{armenia2022participation}. E-participation is a more recent form of citizen involvement using information and communication technologies for electronic government-citizen consultations, thus promoting engagement and collaboration in decision-making \parencite{saebo2008shape,macintosh2004characterizing,alarabiat2021determinants}. It strengthens transparency, advances social inclusion, and fosters democratic societies \parencite{alarabiat2021determinants}. However, implementing e-participation is challenging due to low citizen adoption \parencite{venkatesh2000longitudinal,hassan2020gameful}. Digital literacy gaps, restricted internet access, and poor advertisement can hinder its effectiveness and impact \parencite{alarabiat2021determinants, toots2019participation}.

EU member states have developed a tradition of digital citizen dialogue to reinforce public engagement and active participation in democratic governance \parencite{mariani2024overview}. Policymakers at the local, regional, national, and European levels further increased e-participation efforts in line with the 2020 Digital Agenda as part of the Europe 2020 strategy: \citeauthor*{panopoulou2009eparticipation} (2009) identified 255 different e-participation initiatives within the EU with varying scopes. E-participation employs a variety of tools and platforms to engage citizens, including social media channels like Facebook, YouTube,  and X (ex-Twitter). 
Additionally, dedicated online portals, including bespoke tools such as 
\textit{Have Your Say} were created by governmental agencies to drive citizen involvement. 
Given that legitimate e-participation data is received, there is fragmented knowledge on how to analyze it, particularly for large datasets. For example, \citeauthor*{aitamurto2016civic} (2016) used concept extraction and sentiment analysis on crowd-sourced policy data to evaluate how citizen input shaped a city's transportation plan, while \citeauthor*{hagen2015introducing} (2015) applied textual analysis and topic modeling to classify e-petition texts, thereby exploring policy impacts of emerging themes. Similarly, \citeauthor*{arana2021citizen} (2021) combined topic modeling and text summarization in a direct democracy platform to address information overload and support participation. However, all of these systems have in common that they require prior knowledge of Machine Learning (ML) tools: A gap that LLMs can bridge merely with natural language.

\subsection{Media Richness Theory}
The MRT, initially developed by \citeauthor*{daft1984information} (1984; 1986), provides a theoretical framework for understanding how entities, including public and private organizations and individuals, select communication media to process information effectively. It posits that communication channels vary in capacity to reduce equivocality and manage uncertainty. These early works show that successful entities carefully match the richness of their communication methods with the complexity and uncertainty of the messages they need to share. Critically, they highlighted that organizational success depends on balancing the richness of the media used with the complexity and ambiguity of the tasks, emphasizing that a core challenge often stems from a lack of clarity rather than insufficient data \parencite{daft1984information}. The MRT has been applied to the communication channels governments offer, such as face-to-face, telephone, messenger services, websites, or social media \parencite{sanina2017effectiveness}. Although face-to-face interaction allows for a high richness, it requires a high cost for governments to operate \parencite{androutsopoulou2019transforming}. 

Building on these existing communication channels, the concept of "conversational government" \parencite{baldauf2020towards} offers a solution. It combines the high level of richness found in face-to-face interactions with the cost-effectiveness of automated systems. Utilizing chatbots, governments can thus provide real-time, personalized responses to citizens while lowering costs \parencite{androutsopoulou2019transforming}. Further, there are several examples of how chatbots aligned with MRT principles can advance government-citizen interaction beyond simply providing static information, enabling more dynamic, context-sensitive, and user-focused exchanges. For instance, building on the foundations of automated conversation and enhanced data accessibility, \citeauthor*{cortes2023trends} (2023) developed a holistic framework for employing chatbots in public administration, demonstrating how chatbots can enrich information retrieval, promote greater transparency, and broaden the scope of citizen engagement with governmental services. Similarly, \citeauthor*{segura2022conversational} (2022) illustrate how chatbots facilitate meaningful deliberation by extracting and summarizing complex policy debates, thereby assisting citizens in navigating large-scale civic discussions. Last, LLMs have been utilized to analyze community data, offering insight into refugee needs and supporting data-driven decision-making, thus voicing the needs of refugees and including them in citizen participation \parencite{sprenkamp2025data}. Collectively, these works highlight the potential of chatbots to deliver rich, context-aware communication channels that directly reflect MRT’s emphasis on managing complexity and ambiguity through media choices, ultimately contributing to more responsive, inclusive, and effective governance.

\subsection{Large Language Models}

Artificial Intelligence (AI) has seen significant advancements. In Natural Language Processing (NLP), early static embeddings like GloVe \parencite{pennington-etal-2014-glove} were replaced by context-aware methods following the introduction of the Transformer architecture \parencite{10.5555/3295222.3295349}, which employs self-attention for parallel sequence processing. 
%Models like BERT \parencite{devlin-etal-2019-bert} leverage this architecture to enable tasks such as translation, summarization, and question-answering. 
Generative models further extended NLP capabilities through pre-training on large corpora followed by fine-tuning \parencite{radford2018improving}, culminating in LLMs like GPT-4 \parencite{openai2024gpt4technicalreport}, which integrate multi-modal processing and reinforcement learning for alignment with human preferences. 

Retrieval-Augmented Generation (RAG) methods enhance LLMs by integrating external knowledge. In a typical RAG workflow, the input query is first converted into a vector embedding using an encoder, and all data in the external database are similarly embedded and stored for efficient retrieval. When a query is made, its embedding is used to search the database for the most relevant embeddings, and the retrieved information is then combined with the LLM to generate a more accurate and context-rich response \parencite{lewis2020retrieval}.

LLMs and NLP techniques have been applied in various e-participation contexts to enhance public engagement, streamline service delivery, and support decision-making processes. In citizen input analysis, AI-based systems have automated feedback processing, reducing manual workload while ensuring consistent and scalable results \parencite{borchers2024designing}. In public administration, NLP tools have facilitated policy-making by bridging communication gaps between policymakers and citizens \parencite{guridi2024thoughtful}. Similarly, LLM-powered chatbots have been employed in public services to improve service accessibility, though challenges linked to privacy, transparency, and trust remain critical concerns \parencite{dreyling2024challenges}. Moreover, citizen complaint management systems using contextual feedback mechanisms driven by LLMs have enhanced user trust and engagement by providing meaningful, real-time responses \parencite{10.1145/3640794.3665562}. These applications demonstrate the practical potential of LLMs in enabling more interactive, transparent, and efficient e-participation systems over previous ML-based implementations. However, this novel approach is yet to be applied to a real-world context.

\section{Methodology}
This study is part of a larger project focused on improving EU citizen participation through an LLM-enabled chatbot. As such, enabling policymakers to effectively analyze e-participation data is a necessary first step to drive more active citizen participation. To create a solution for this real-world problem \parencite{nunamaker2015last}, we employ the DSR methodology \parencite{hevner2004design, peffers2007design}, namely, 1) problem identification, 2) definition of solution objectives, 3) design and development, 4) demonstration, 5) evaluation, and 6) communication.

To (1) identify the problems related to EU citizen participation and (2) derive solution objectives, we observed the \textit{Have Your Say} platform and compared it to insights from academic literature. These findings informed the definition of our problem statements and solution objectives, which aimed to address these specific challenges through the development of an advanced chatbot. For the (3) design and development phase, we iteratively created our prototype—a web-based LLM-powered chatbot, called \textit{AskThePublic}\footnote{\url{http://askthepublic.eu/}}, the source code can be found on GitHub\footnote{https://github.com/HICSS26/AskThePublic}. It is built with React for the frontend and FastAPI for the backend, uses LangChain for NLP tasks, interacts with MongoDB and OpenAI, and is containerized with Docker to support reproducibility and streamlined deployment.

%add here the description about the government collaboration
Next, it is essential to (4) demonstrate and (5) evaluate the artifacts created to determine if the research objectives have been met. For the demonstration and evaluation, we conducted 11 semi-structured interviews (ID1-ID11). In order to be eligible, participants needed to be EU citizen who were affiliated with the local government or EU institutions through research projects. We gathered participants from five countries (France, Germany, Greece, Italy, and Luxembourg) and varying backgrounds, including law (8,10), finance (ID7,9), security and privacy (ID3-6), and energy (ID1-2,11). The participants had no previous work experience related to citizen participation or the \textit{Have Your Say} platform. Prior to the session, a short survey was administered to assess the participants' perceptions of the EU. During the sessions, participants interacted with the chatbot, performing tasks such as querying content on specific EU initiatives, focusing on the usability and effectiveness of the solution within their field of expertise, as well as informing directions for future development based on their professional assessment. The sessions, ranging between 44 and 74 minutes, have been recorded, transcribed, and analyzed using a qualitative approach \parencite{saldana2021coding} consistent with Gioia's methodology for inductive research \parencite{gioia2013seeking}. The session guideline is available upon request. Finally, the (6) communication of this research is achieved through the present article.

\section{Results}
\subsection{Problem Statement \& Solution Objectives}
\label{subsec:problem_solution}

The \textit{Have Your Say} platform provides several opportunities for broad citizen participation. Yet, it faces several limitations that can hinder its effectiveness. We found three major problems in analyzing data within the \textit{Have Your Say} platform, presented in the following. Table \ref{table:problem_solution_artifact} matches those with the solution objectives, and offers an overview of which participants provided a favorable evaluation, with the remaining ones provided either neutral, or critical feedback of the solution.

\textbf{Lack of Interaction} constitutes a major issue, as the platform primarily facilitates one-way communication. Citizens can submit feedback, but the lack of real-time responses or interactive features limits their involvement, aligning with the lower levels of the public participation spectrum, namely informing, consulting, and involving rather than collaborating or empowering \parencite{shipley2012making}. In particular, when users interact with the platform, they must first navigate to an initiative of interest, which can be challenging since most EU citizens are unclear about the objectives and context of various initiatives. After selecting an initiative, users are expected to navigate the timeline and supporting documentation, often consisting of text-heavy external files. The feedback submitted on the platform is ordered by time of submission, while lacking filtering or clustering options. The proposed solution should address this by providing interactive functions for more effective communication (\citeauthor*{daft1986organizational} 1984; 1986).
$
$

\begin{table}[h!]
\centering
\footnotesize	
\caption{Overview of Selected DSR Steps}
\begin{tabularx}{\linewidth}{|p{32mm}|p{20mm}|X|}
\hline
\textbf{Problem Statements} & \textbf{Solution Objectives} & \textbf{Positive Evaluation} \\ \hline

\textit{Lack of Interaction} (Daft and Lengel, 1984; 1986; \citeauthor*{shipley2012making}, 2012) & 
Real-Time Interaction with Citizen Feedback & 
ID 1-3, 6-11 \\ \hline

\textit{Lack of Structure} \parencite{arana2021citizen, lourencco2007incorporating, armenia2022participation} & 
1) Structured Feedback 

2) Stakeholder Clustering & 
1) ID 1, 3-4, 9-11  

2) ID 1-5, 7-8, 10-11 \\ \hline

\textit{Lack of Understanding} \parencite{wright2009elephant, alarabiat2021determinants, toots2019participation} & 
Automated Multilingual Synthesis & 
ID 1, 10-11 \\ \hline

\end{tabularx}
\label{table:problem_solution_artifact}
\end{table}

\textbf{Lack of Structure} is another major issue in feedback collection due to the use of natural language. Responses vary in length, focus, and clarity, making it difficult for stakeholders to extract meaningful insights without automated tools to categorize and filter input \parencite{arana2021citizen}. In particular, to polarizing topics, over 1000 feedback statements are submitted, requiring the user to distinguish qualified feedback from unqualified one, particularly from anonymous platform users. Frequently, feedback does not relate to the topic of the initiative or merely scratches the surface, making it difficult for EU citizens, but also for policymakers to fully comprehend the feedback and to leverage it as part of the implementation process. The proposed solution is expected to offer tools that make the structuring of the feedback by stakeholder groups more efficient \parencite{lourencco2007incorporating, armenia2022participation}.
$
$

\begin{figure*}[h!]
    \centering
    \includegraphics[width=0.65\linewidth]{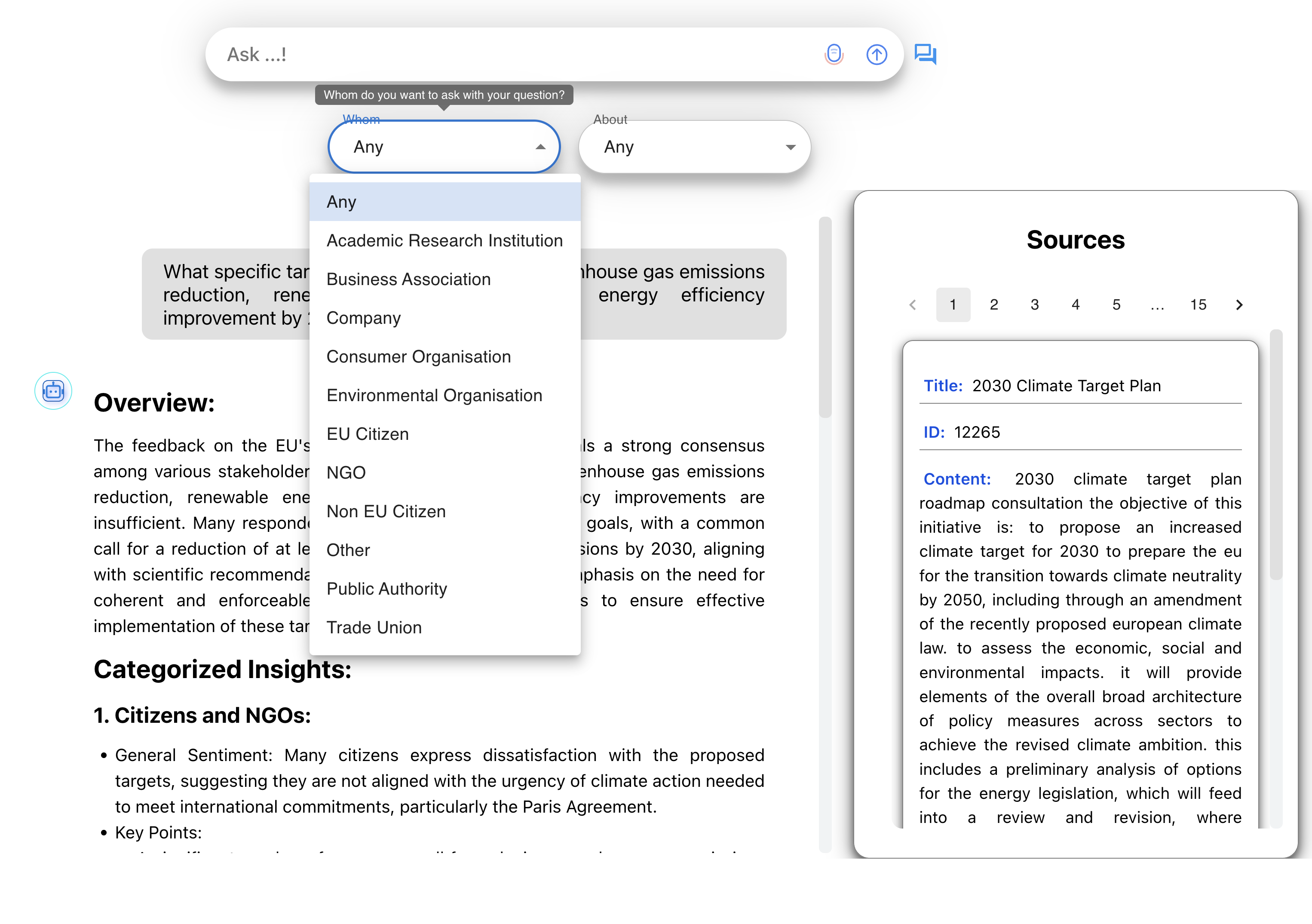}
    \caption{Analysis page displaying a detailed response to the user's query}
    \label{fig:askthepublic_ans}
\end{figure*}

\textbf{Lack of Understanding} is a third issue stemming from the diverse linguistic and cultural landscape of the EU. As feedback may be written in any of the 24 official languages of the EU, without providing tools for integrated translation, barriers to understanding and accessibility arise for both citizens and policymakers \parencite{wright2009elephant}. In particular, certain initiatives may receive mainstream media attention in selected countries. In those circumstances, it is common to find that citizens of those countries submit most of the available feedback statements. Thus, the proposed solution is expected to offer output comprehensible by all EU citizens \parencite{alarabiat2021determinants, toots2019participation}. 
$
$

\subsection{Artifact Design}

To address the problems mentioned in Section \ref{subsec:problem_solution}, we developed \textit{AskThePublic} (Figure \ref{fig:askthepublic_ans}). It provides an intuitive interface for user interaction, and at its core, it utilizes RAG to answer user queries effectively.

\begin{figure}[h]
    \centering
    \includegraphics[width=0.8\linewidth]{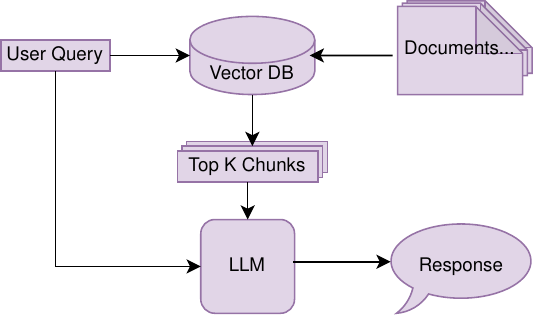}
    \caption{The RAG pipeline}
    \label{fig:pr}
\end{figure}

The interface is designed to be accessible and straightforward. It contains a search bar that allows users to input queries in natural language. This central element is complemented by two drop-down filters: ``Whom'' and ``About''. These filters enable users to narrow their queries by selecting specific groups they want to inquire about and the topics they are interested in exploring further. By default, the system searches across all defined groups and all stored comments within various topics to provide comprehensive responses. 

RAG is the core framework driving the system, combining advanced information retrieval techniques with the generative capabilities of LLMs. We rely on RAG because even models like GPT-4 can produce inaccurate information, and RAG has proven effective in mitigating this issue. To support this, the system starts by scraping feedback data from the \textit{Have Your Say} platform. This feedback, often unstructured and multilingual, is preprocessed to remove noise and inconsistencies. The data is then transformed into dense vector embeddings using  OpenAI’s text-embedding-3-small%\footnote{\url{https://platform.openai.com/docs/guides/embeddings}}
, which encode the semantic meaning of the text into a high-dimensional numerical representation. These embeddings are stored in a MongoDB Atlas vector database, enabling efficient semantic search during retrieval.

When a user submits a query, the system first processes it by generating a vector embedding for the input, matching it against the database of stored feedback using cosine similarity. The system then returns the $K$ entries from the database with the highest cosine similarity as the context for LLM. At this stage, the system leverages GPT-4o-mini by feeding the retrieved context along with the user’s query into the LLM. This integration delivers reliable, insightful summaries and actionable recommendations in an intuitive, human-like format. The RAG pipeline process to generate final response is shown in Figure \ref{fig:pr}.

%Additionally, a microphone icon allows users to input queries via voice, enhancing accessibility and ease of use. The ``Ask'' button initiates the query process, leading users to an analysis page that displays summaries and categorized insights derived from the data. 

The output of the chatbot is structured into three parts. The first part provides an overview, highlighting key themes and sentiments from public opinions. The second part offers categorized insights, showcasing feedback segmented by different user groups and summarizing key points for each group. Finally, the third part delivers actionable insights, presenting two to three specific, evidence-based recommendations or highlighting conflicts that require resolution. To increase user confidence and transparency, the system includes a Source section, which highlights the resources used to generate the answer to the given query. A new chat icon positioned beside the search bar enables users to refresh the page and initiate a new query.

\subsection{Demonstration and Evaluation}
A total of 357 in-vivo codes were derived during the analysis of the 11 semi-structured interviews. These were then clustered into the solution objective categories, namely, real-time interaction (85 in-vivo codes), structured feedback management (69 in-vivo codes), and multilingual synthesis (63 in-vivo codes). A fourth category, prospective stakeholder usage, was introduced to inform future research directions, covering the remaining 140 in-vivo codes. The analysis of the pre-study shows that the participants shared a rather positive perception of the EU, with an average of 8.0 on a scale from 1 (very negative) to 10 (very positive). In the following, the results of the qualitative assessment according to the solution objectives are presented.

\textbf{Real-Time Interaction} Although some participants (ID4-6) were sceptical whether the RAG capabilities provide enough benefit over established chatbots (\textit{“Everything sounds reasonable. I probably would need to ask the same question to ChatGPT, because, based on what I read, though it is neither right nor wrong, I do not see why we need another chatbot.”, ID5}), the majority of the participants (ID 1-2,7-9) indicated they are more likely to use \textit{AskThePublic} over the \textit{Have Your Say} platform due to the enhanced interaction: \textit{“I would be much more likely to use it than to browse through every single initiative or directive and to read the comments separately, this is obviously made much easier. And it also reduces the entry barrier, so to say, to inform myself. (ID8)”} One participant narrowed the benefit of the solution down to a single statement: \textit{“Basically, the more advanced search engine for public opinions. (ID4)”} Participants have particularly praised its intuitiveness: \textit{``It’s easy. In the beginning, you are doing a test. If you are not sure, then you see how it works. (ID9)''} The tool was further noted to be well suited for information purposes (ID 2-3, 6-10), particularly in cases where one does not have pre-existing knowledge of a domain: \textit{“So, if you do not have a very precise question in mind, but you rather want to assess opinion in certain areas of topics, then, it is being summarized really well. […] I really like how it works because you can receive various opinions in a fairly well-structured and simple way. (ID3)”} Moreover, several participants (ID 1-2,7,10-11) noted the various features of the chatbot positively: \textit{“This is something I really enjoy, that I have my text on the left side, and on the right side the sources for validation. (ID10)}”

\textbf{Structured Feedback Management} Participants generally view the structure favorably (ID 1,3-4,9-11), as exemplified by ID10: \textit{“I actually like that it gives a broad answer at the top, then going into more insights and also having an implication section. So it has this tunneling effect.”} The structure allows its users to conduct further research: \textit{“It is not too long, not too short, so it provides you an overview and from there you can take it further and research yourself. (ID9)”} Participants particularly highlighted the ability to engage with specific stakeholder groups directly (ID1-5, 7-8, 10-11): \textit{“If you ask specific groups of people questions, it seems that you can engage with these people. You can engage with the answers that they have already provided on the platform. (ID7)”} At the same time, through RAG-based computation, \textit{AskThePublic} only retrieves stakeholder feedback that relates directly to the query, in stark contrast to the \textit{Have Your Say} platform. While this was noted positively, all participants questioned how the sources are being selected (ID1-11): \textit{“If you say, 1000 people have responded, but now I get four random universities as output, I question the integrity of the process, what makes them stand out (ID6)?”} Participants noted an apparent lack of credibility when it comes to certain stakeholders that are being displayed: \textit{“Perhaps the German AI Association I have heard before, but I do not know any other, so I cannot judge if that is a source where perhaps three people are sitting who simply do not have any relevance, or if it is some sort of NGO which is truly important. (ID3)”} One participant suspected a potential bias towards more polarizing opinions that are selected to be displayed: \textit{“So for example, not everything about dynamic tariffs is bad, but it feels like this kind of overview right now focuses pretty heavily on the bad side. (ID11)”} When being asked very specific questions, the chatbot was found to create answers artificially. This is recognized as a major threat to the validity: \textit{“So if you go and search yourself, and you find out that there is nothing, this is very bad for the reputation of the system. (ID9)”}

\textbf{Multilingual Synthesis} The solution translates stakeholder feedback given in any official EU language to the language the question was asked in, making it easier for EU citizens to understand the feedback (ID1, 10-11), as exemplified by ID 10: \textit{“The generation of my grandparents, or even my parents, they do not speak English so well. So for them, this is valuable because if they are struggling to understand the text, they will go to some tabloid newspaper, even if they do not provide accurate information.”} However, it was noted that this only applies to the immediate text output from the chatbot and does not apply to the sources panel. Further, finding the right sourcing balance is commonly criticized (ID4-5, 10-11). On the one hand, participants found larger countries with bigger impact most relevant, as exemplified by ID 4 (\textit{“In this question, I only got answers from Germany and Slovakia. Germany, I understand, but as for Slovakia, frankly, I do not understand why it is being displayed”}), on the other hand, participants noted that they would like to have a bigger variety, as exemplified by ID 10 (\textit{“I would prefer to have three bullet points from three different countries instead of having the majority from one country.”})

\section{Discussion}
Utilizing DSR, we are able to answer our RQ. We do so by discussing the solution objectives and design features of the designed artifact \textit{AskThePublic}  (Section \ref{sec:discussion_designed_artifact}). Further, we give insights on how different stakeholders can utilize the given design (Section \ref{sec:discussion_usergroups}).
\subsection{Designed Artifact and Future Research}
\label{sec:discussion_designed_artifact}
The evaluation of the designed artifact demonstrates that each of our three solution objectives holds significant promise for enhancing e-participation.

Participants appreciated the chatbot's intuitiveness and ease of use, noting that its real-time interaction capabilities significantly enhanced user engagement compared to traditional platforms. This finding aligns with the MRT, which posits that richer media facilitate more effective communication (\parencite{daft1984information}. The chatbot’s ability to provide immediate, interactive responses mirrors the high richness of face-to-face communication, thereby increasing its effectiveness in engaging users as a practical example of conversational government \parencite{baldauf2020towards}.

\textit{AskThePublic's} ability to organize unstructured feedback into coherent and actionable insights was another key strength noted by participants. However, in the absence of relevant information to display, the chatbot might hallucinate, thus leading some participants to question the added benefits over LLMs without RAG, underlining that the system is only as good as the available data. Provided there is relevant data, the system effectively filters and structures feedback, ensuring that only relevant information is presented \parencite{lewis2020retrieval}, thus focusing on transparency, a quality needed in e-participation systems \parencite{alarabiat2021determinants}. While this concept was rooted in \citeauthor*{sprenkamp2025data} (2025), who analyzed social media data to make it available for policy-making, \textit{AskThePublic} allows for generating insights from e-participation data in natural language, a large contribution over prior ML-based tools \parencite{armenia2022participation,hassan2020gameful,mariani2024overview}.As participants found this structured format facilitated easier navigation and deeper understanding of the e-participation data, we foresee that systems like \textit{AskThePublic} will attract higher adaptation rates, thus addressing a common problem among modern e-participation tools \parencite{venkatesh2000longitudinal, hassan2020gameful}.

The multilingual capabilities of \textit{AskThePublic} helped make feedback accessible across the EU's 24 official languages. This feature directly addresses language barriers identified as significant determinants affecting e-participation \parencite{alarabiat2021determinants}. By utilizing LLMs for translation and synthesis, the chatbot fosters a more inclusive environment, aligning with the principles of conversational government that advocate for language-inclusive communication channels \parencite{baldauf2020towards}. Further, while the chatbot answers in any of the official EU languages, it further synthesizes context from various source documents in the RAG-system \parencite{sprenkamp2025data}. Concerns regarding the balance and diversity of sources were raised, indicating that the chatbot should implement strategies to ensure a more equitable distribution of feedback from various countries. Notably, the current embedding fetching mechanism prioritizes data points in the language of the query, which may inadvertently limit the diversity of retrieved multilingual feedback, highlighting the need to balance language prioritization within the retrieval process. These limitations highlight the need for comprehensive multilingual support to further inclusion.

In addition, users proposed new features for \textit{AskThePublic}. To mitigate the bias towards polarizing opinions and ensure a more balanced representation of feedback, the system could incorporate algorithms that promote diversity in source selection, aligning with research on unbiased data representation in public administration \parencite{alarabiat2021determinants}. Another option is to utilize custom models based on open-source projects. Currently, we utilize the GPT API, a closed-source model. Thus, the training data and potential biases cannot be identified; here, open-source LLMs like EURO-LLM \parencite{martins2024eurollm} could be leveraged to align more closely with EU values of transparency and data sovereignty. While open-source models often lack the accuracy of closed-source models, this has been recently challenged by the development of DeepSeek \parencite{guo2025deepseek}, which exceeds the efficiency of state-of-the-art LLMs while being on par in accuracy. Thus, we expect governments to be empowered in the future to utilize open-source solutions. Several users flagged that \textit{AskThePublic} could integrate unique functionalities not offered by existing chatbots like ChatGPT, such as specialized data visualization tools or tailored stakeholder engagement features, enhancing its distinct value proposition \parencite{borchers2024designing}. Furthermore, expanding the chatbot’s integration capabilities with other e-participation platforms and social media channels can broaden its accessibility and usability, as highlighted in related works on digital engagement strategies \parencite{macintosh2004characterizing}. By implementing these recommendations, future iterations of \textit{AskThePublic} can enhance its effectiveness, inclusivity, and trust, thereby further supporting democratic engagement.

Last, in line with MRT \parencite{daft1984information,daft1986organizational}, communication media that allow immediate feedback and multiple cues are more effective at reducing equivocality. AskThePublic’s chat interface, by offering real-time, two-way dialogue and transparent sourcing, functions as a richer medium compared to the static Have Your Say portal. This increased richness directly maps to participants’ reports of greater engagement and clearer understanding.

\subsection{Possible User Groups}
\label{sec:discussion_usergroups}
\textbf{EU policymakers} can leverage \textit{AskThePublic} to gather and analyze feedback more effectively, thus informing more inclusive policy decisions, which are based on public data. By incorporating features that balance data distribution across different demographics, the chatbot can enhance both the quality and reliability of the feedback used in policy-making, addressing challenges related to sampling bias and data integrity \parencite{alarabiat2021determinants}.

\textbf{For politically engaged EU citizens}, \textit{AskThePublic} holds significant potential by providing an intuitive and interactive platform for accessing and analyzing policy feedback. Participants further valued the empowerment and active participation the tool could foster (ID8, ID6, ID10). By making feedback more accessible and structured, the chatbot can enhance citizen engagement and address common barriers to e-participation, such as lack of information and perceived impact.

\textbf{For journalists and researchers}, \textit{AskThePublic} provides a valuable tool for extracting feedback on legislation. Thus, they can identify the sentiments of citizens or organizations towards a given topic, providing researchers and journalists with a reliable and unique source that can be used for publications. However, future iterations of \textit{AskThePublic} need to design clear links to original feedback that can support accurate reporting and data-driven research.

\section{Conclusion}
While e-participation has been in the focus of the academic discourse \parencite{armenia2022participation,hassan2020gameful,mariani2024overview}, methods to analyze the submitted feedback are sparse and focusing on the fragmented usage of ML methods to classify citizen needs \parencite{hagen2015introducing,aitamurto2016civic,arana2021citizen}. The advent of LLMs enables novel applications that utilize NLP to analyze citizen feedback, needing little to no understanding of the technology, as shown through \textit{AskThePublic}.

Using the DSR methodology \parencite{hevner2004design,peffers2007design}, the development of \textit{AskThePublic} has made three contributions. First, we provide an initial tool for the analysis of citizen feedback using LLMs, which are open to the general public, enabling users to ask questions without the need for expertise in ML or computational tools. Second, we give design recommendations for tools enabling the structured analysis of e-participation. We hope that these recommendations are useful for further development in academia and the public sector. Third, we identify the first user groups for our tool. 

However, our study is not without limitations. Our evaluation was solely done with a rather small user group with largely positive views of the EU, leaving room for broader evaluations. Moreover, we developed and tested \textit{AskThePublic} solely using the GPT-4 API, creating a bias towards the training data of this closed-source model. We see these limitations as opportunities for future work. Later, iterations of \textit{AskThePublic} or similar tools should further evaluate our design recommendations with the goal of generating long-lasting design principles \parencite{gregor2020research}. Moreover, the effect of different closed and open-source LLMs should be tested to align with the values of the EU, e.g., EURO-LLM \parencite{martins2024eurollm}. \\

\noindent\textbf{Acknowledgements} This research was funded in part by the Luxembourg National Research Fund (FNR) and PayPal,
PEARL grant reference 13342933/Gilbert Fridgen. We also thank the University of Zurich and the Digital Society Initiative for (partially) financing the study through Liudmila Zavolokina's DIZH postdoc fellowship. For the purpose of open access, and in fulfillment of the obligations arising from the grant agreement, the author has applied a Creative Commons Attribution 4.0 International (CC BY 4.0) license to any Author Accepted Manuscript version arising from
this submission.

% Fonts specification --- not shown as it doesn't exist in the Word document either. 

%\section{Fonts}

%A summary of fonts is provided in Table \ref{tab: fonts}. 

%\begin{table}[thb]
%\centering
%\caption{\label{font-table} Font guide. \vskip 3pt }
%\label{tab: fonts}
%\begin{tabular}{l|rl}
%\hline \bf Type of Text & \bf Font Size & \bf Style \\ \hline
%paper title & 14 pt &  \bf bold \\
%authors & 10 pt &  \underline{email} underlined \\
%abstract title & 12 pt &  \bf bold\\
%abstract text & 10 pt &  \it italic\\
%section titles & 12 pt & \bf bold \\
%subsection titles & 11 pt & \bf bold \\
%document text & 10 pt  & \\
%captions & 9 pt & \sansserifformat{\captionsize sans-serif, \bf bold} \\
%bibliography & 9 pt & \\
%footnotes & 8 pt & \\
%\hline
%\end{tabular}
%\end{table}

% if added before the last page, this command can help balancing columns
%\addtolength{\textheight}{-.2cm} 

%Bibliography 
% \bibliographystyle{apalike}
% \bibliography{sample}

% \begin{figure*}%[htbp]
%   \centering
%   \begin{minipage}[b]{0.695\textwidth}
%     \centering
%     \includegraphics[width=\textwidth]{images/askthepublic_ans.png}
%     \caption{Caption for Figure 1}
%     \label{fig:figure1}
%   \end{minipage}
%   \hfill
%   \begin{minipage}[b]{0.295\textwidth}
%     \centering
%     \includegraphics[width=\textwidth]{images/rag.pdf}
%     \caption{Caption for Figure 2}
%     \label{fig:figure2}
%   \end{minipage}
% \end{figure*}

\printbibliography
\end{document}